\documentclass[10pt]{article}
\usepackage{epsf}
\usepackage{latexsym}
\newcommand{\nix}[1]{}
\title{\large \bf Spin-photocurrent in $p$-SiGe quantum wells  under terahertz laser irradiation}
\author{V.~V.~Bel'kov\,$^{1,2}$, S.~D. Ganichev\,$^{1}$, Petra Schneider\,$^1$,
D.~Schowalter$^1$, \\ U.~R\"{o}ssler\,$^1$, W.~Prettl\,$^1$, E.~L.~Ivchenko\,$^2$, R.~Neumann\,$^3$, K.~Brunner\,$^3$ and G.~Abstreiter\,$^3$}

\date{\normalsize
$^1$ Fakult\"{a}t f\"{u}r Physik, Universit\"{a}t Regensburg, D-93040 Regensburg,
Germany \\
$^2$ A.F.~Ioffe Physico-Technical Institute, 194021 St. Petersburg,
Russia\\
$^3$\,Walter Schottky Institute, TU Munich,  D-85748 Garching, Germany
}

\begin{document}
\maketitle
\thispagestyle{empty}
{\bf Abstract:}
A  detailed study of the circular photogalvanic effect (CPGE) in SiGe structures is presented.
It is shown that the CPGE becomes possible
due to the built-in asymmetry of quantum wells (QWs) in compositionally stepped samples
and in asymmetrically doped structures.
The photocurrent arises due to optical spin
orientation of free carriers in QWs with  spin  splitting in
{\boldmath$k$}-space.
It is shown that the effect can be applied  to probe the
macroscopic
in-plane symmetry of low dimensional structures and allowing to conclude on
Rashba or Dresselhaus terms in the Hamiltonian.
\\
\subsection*{\centerline{Introduction}}
Recently it has been demonstrated that in quantum well structures based on
III-V compounds an electric  current linked to spin-polarized
carriers can be generated  by circularly polarized light. Two such effects
were observed: the circular photogalvanic effect~\cite{PRL01} and
the spin-galvanic effect~\cite{Nature02}.
Microscopically both effects are based on {\boldmath$k$}-linear terms in the
electron Hamiltonian known as
Rashba and Dresselhaus terms~\cite{Rashba,Dyakonov} which lift the spin degeneracy of electron subbands.
The current flow is driven by an
asymmetric distribution of carriers in the spin-split subbands.

On a phenomenological level  a current due to a spin polarization as well as {\boldmath$k$}-linear terms in the band structure become possible in systems belonging to one of the gyrotropic crystal classes~\cite{Cardona}. This condition is met by zinc-blende-structure based QWs.
In materials of T$_d$-symmetry which lack a center of inversion,
gyrotropy is obtained due to the reduction of the dimensionality alone.
In
contrast, in low dimensional structures based on Si and Ge (SiGe QW) which
have  inversion symmetry both effects are forbidden by symmetry.
Recently we
have shown that even in such structures spin photocurrents may be obtained if the  inversion symmetry is broken by preparation of compositionally stepped
quantum wells or asymmetric doping of compositionally symmetric quantum
wells~\cite{PRB02}. Here we present a detailed study of the CPGE and
demonstrate that investigation of the photogalvanic current with respect to the
crystallographic directions allows to determine the macroscopic in-plane
symmetry of QW structures.
\subsection*{\centerline{Theoretical consideration}}
The principal microscopic aspect of a photon helicity driven spin
photocurrents like CPGE is a removal of spin-degeneracy in the subband
states due to the reduced symmetry of the quantum well structure~\cite{PRL01,Nature02}. It
is related to the appearance of {\boldmath{$k$}}-linear terms in the
Hamiltonian,
\begin{equation}
H^{(1)}(\mbox{{\boldmath$k$}})\,=\,\sum_{lm}\beta_{lm}\sigma_l k_m
\end{equation}
where {\boldmath$\beta$} is a pseudotensor
and the {\boldmath$\sigma$} are the
Pauli spin matrices.
As discussed in~\cite{PRL01} the coupling between
the carrier spin ($\sigma_l$) and momentum ($k_m$) together with the
spin-controlled dipole selection rules yields a net current under circularly
polarized excitation. Depending on the photon energy this spin photocurrent can
be either due to direct or indirect intraband transitions.

Spin degeneracy results from
the simultaneous presence of time-reversal and spatial inversion symmetry.
If one of these symmetries is broken the spin degeneracy is lifted.
In our SiGe QW systems the spatial inversion symmetry is broken
and, as a consequence, spin-dependent {\boldmath$k$}-linear terms appearing in the
electron Hamiltonian lead to a splitting of the electronic subbands at non-zero
in-plane wavevector.
In the context of spin related phenomena in QW structures most frequently
the  Rashba term of the  the form $\sigma_x k_y -
\sigma_y k_x$ is taken into account being  caused by a structural inversion asymmetry
(SIA)~\cite{Wolf}.

In zinc-blende-structure based QWs
a {\boldmath$k$}-linear term
proportional to
$\sigma_x k_x - \sigma_y k_y$ (Dresselhaus term) is also present due to
the  bulk inversion asymmetry (BIA). We have shown in~\cite{PRB02}
that a term of this form can also be obtained in  (001)-grown SiGe
structures of C$_{2v}$ symmetry, which does not have bulk inversion
asymmetry. In this case the tetrahedral orientation of  chemical bonds
at interfaces  gives rise to a coupling between heavy and light hole
states~\cite{xx13}. We would like to emphasize that Dresselhaus terms of this
type may also be present in zinc-blend-structure based QWs that have not
yet been considered. Several
microscopically different mechanisms leading to {\boldmath$k$}-linear terms
of both types were discussed in~\cite{PRB02}.

The phenomenological theory of the CPGE allows us to probe  the
macroscopic in-plane symmetry of QW structures. The CPGE is
described by
\begin{equation}
\label{eq1}
j_{\lambda} = \sum_{\mu}\gamma_{\lambda \mu}\:i (\mbox{{\boldmath$E$}} \times
\mbox{{\boldmath$E$}}^*
)_{\mu} \:,
\end{equation}
where {\boldmath$j$} is the  photocurrent density,
{\boldmath$\gamma$} is a second rank pseudotensor,
{\boldmath$E$} is the
complex amplitude of the
electric field of the electromagnetic wave, and
$i (${\boldmath$E$}$ \times ${\boldmath$E$}$^* )_{\mu} = \hat{e}_\mu
P_{circ}\:E_0^2$,
with $E_0$, $P_{circ}$, {\boldmath$\hat{e}$} = {\boldmath$q$}$/q$ and
{\boldmath$q$} being  the
electric field amplitude, the degree of light circular polarization, the unit
vector pointing in
the direction of light propagation, and  the light wavevector inside the medium,
respectively.
The second rank pseudotensors
{\boldmath$\gamma$},  {\boldmath$\beta$}, and the
tensor of gyrotropy are isomorphic.
Nonzero components of these tensors may exist
if at least one component of  a polar and an axial vector transforms
according to the identity
representation of the corresponding point group.

\nix{
In general, in addition to
the photocurrent given in Eq.~(\ref{eq1}),
two other photocurrents can be observed
simultaneously, namely the linear photogalvanic effect (LPGE) described by
$j_{\lambda} = \sum_{\mu\nu}\chi_{\lambda \mu \nu} (E_{\mu} E^*_{\nu} + E_{\nu}
E^*_{\mu})/2$ and the photon drag effect $j_{\lambda} = \sum_{\mu\nu\delta}
T_{\lambda \mu \nu
\delta} E_{\mu} E^*_{\nu} q_{\delta} \:.$ Both effects were observed in
quantum well structures~\cite{PRB02,APL00,photondrag} but they do not
require spin orientation and therefore are outside the scope of the present
investigation.
}
\begin{figure}[tb]
\begin{center}
\mbox{\epsfysize=4.8cm \epsfbox{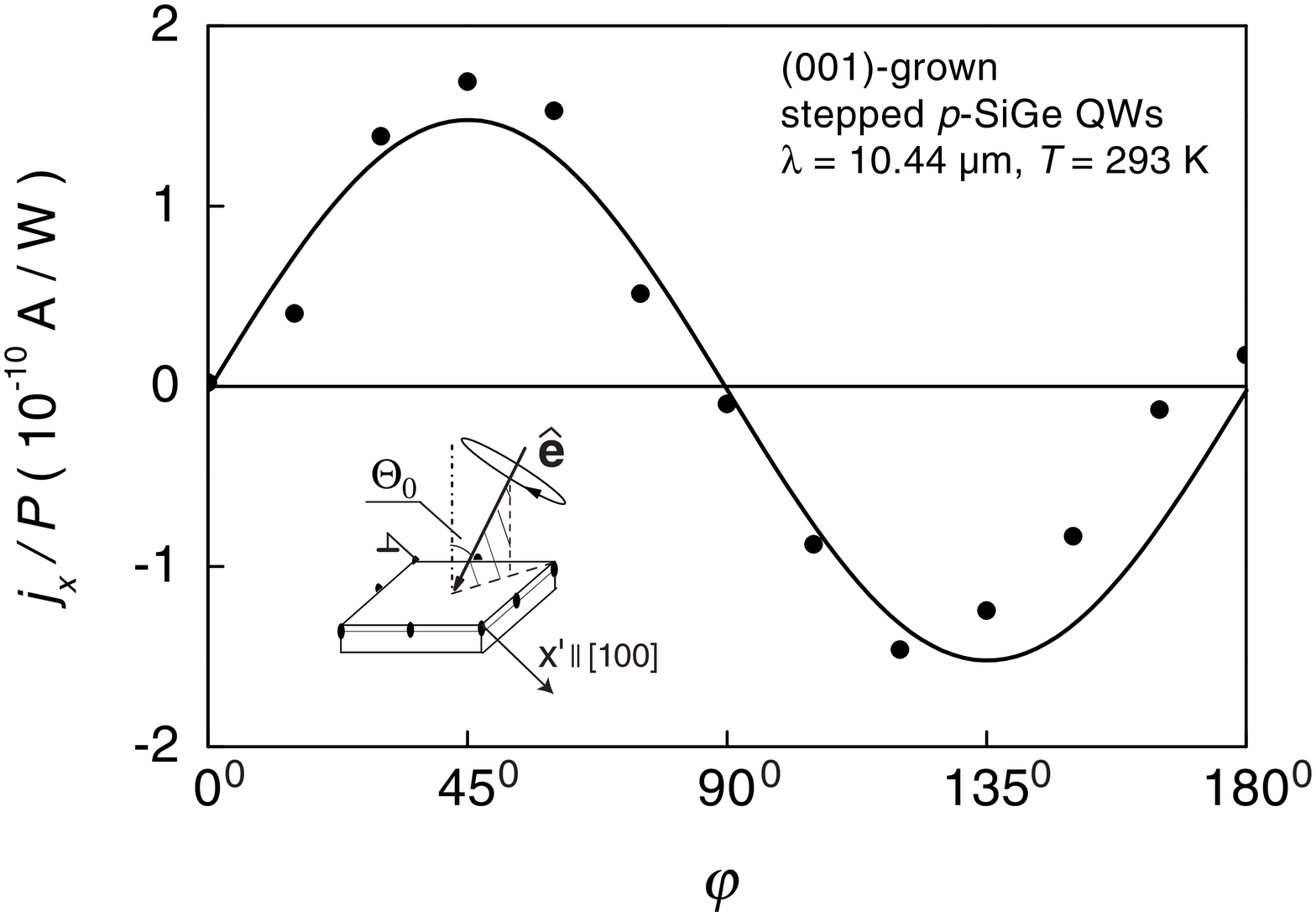}
}
\caption{ Photogalvanic current $j_x$ in (001)-grown
compositionally stepped SiGe QWs normalized by the light power~$P$ measured
at room temperature as a function of the phase angle $\varphi$. The data
were obtained under oblique incidence $\Theta_0$= 30$^\circ$ of irradiation
at $\lambda = 10.44\,\mu$m. The full line is fitted after Eq.~\protect
(\ref{jxjy}). }
\end{center}
\end{figure}

Nonzero components of {\boldmath$\gamma$} for (001)- or (113)-grown SiGe
QWs can be obtained by preparation of compositionally stepped QWs and
asymmetric doping of compositionally symmetric QWs. Several point groups
are relevant in connection with the photogalvanic experiments described
below. For asymmetric QWs (due to doping or different profiles of the left
and right interfaces) the macroscopic symmetry is reduced to C$_{2v}$. If
QWs are grown along the low-symmetry axis $z \parallel [hhl]$ with $[hhl]
\neq$ [001] or [111] the point group becomes C$_s$ and thus CPGE is also
allowed as it is the case for zinc-blende-structure based QWs grown on
(113)-oriented substrates.

Making use of the sample symmetry we derive the photocurrent {\boldmath$j$} of
Eq.~(\ref{eq1}) as a function of the light helicity. First we use
cartesian  coordinates $x, y, z$  along the directions [1$\bar{1}$0], [$ll
(\overline{2h})$] and [$hhl$], respectively,
where [$hhl$] ([001] or [113]
in our case) is the growth axis of the QW structure. Due to carrier
confinement in $z$ direction the photocurrent in QWs has nonvanishing
components only in $x$ and $y$. Then, in a system of C$_{2v}$ symmetry, the
tensor {\boldmath$\gamma$} describing the CPGE has two
linearly independent components $\gamma_{xy}$ and $\gamma_{yx}$ and
Eq.\,(\ref{eq1}) reduces to

\begin{equation} \label{jxjy}
j_{x} = \gamma_{x y} \hat{e}_y  P_{circ} E_0^2\:,\: \: \: \:\:\:\:\:\:
j_{y} = \gamma_{y x} \hat{e}_x  P_{circ} E_0^2.
\end{equation}
The same equations are also valid for the  point group D$_{2d}$
but
this higher symmetry
imposes the condition $\gamma_{xy} = \gamma_{yx}$ on the
{\boldmath$\gamma$} tensor components.

For both symmetries, D$_{2d}$ and C$_{2v}$, a circular photocurrent can
be induced only under
oblique incidence of radiation because for  normal incidence,
$\mathbf{\hat{e}} \parallel [001]$ and hence $\hat{e}_x= \hat{e}_y = 0$.
Thus rewriting the components $\gamma_{\lambda \mu}$ in the form
$\gamma_{x y} = \gamma_{1} + \gamma_{2} $, $\gamma_{y x} = \gamma_{1} -
\gamma_{2} $ and substituting this into Eq. (\ref{jxjy}) we can
consider the coefficient $\gamma_{2}$ as a signature of the
symmetry reduction from D$_{2d}$ to C$_{2v}$.
Choosing an other coordinate system ($x', y', z$) with the directions
parallel to [100], [010] and [001], respectively, we obtain for the circular
photogalvanic current
\begin{equation} \label{jx1jy1} j_{x'} = E_0^2
P_{circ} ( \gamma_{1} \hat{e}_{x'} + \gamma_{2} \hat{e}_{y'}) \:,\: \: \:
\:\:\:\:\:\: j_{y'} = E_0^2 P_{circ} ( -\gamma_{2} \hat{e}_{x'} -
\gamma_{1} \hat{e}_{y'}).
\end{equation}
While for $x, y$ coordinates
only a transverse effect occurs, see  Eq.~(\ref{jxjy}), for $x', y'$ directions,
both longitudinal and transverse effects may be present,
see Eq.~(\ref{jx1jy1}). This fact allows to make use of the CPGE for
investigation of the macroscopic in-plane symmetry of QWs. Indeed, as it has
been shown in~\cite{PhysicaE01_2} for the D$_{2d}$ symmetry group $\gamma_{2}$
is equal to zero and, therefore, no transverse photogalvanic current can
be generated at the excitation by light along $x'$ or $y'$ directions.

In contrast to structures of
C$_{2v}$ symmetry in structures of C$_s$ symmetry the CPGE is allowed
for normal incidence {\boldmath$\hat{e}$}$ \parallel [hhl]$ because in this
case the tensor {\boldmath $\gamma$} has the additional nonzero component
$\gamma_{xz}$. The current flows along [1$\bar{1}$0] direction,
perpendicular to the mirror reflection plane, and is described by

\begin{equation} \label{113}
j_{x} = \gamma_{xz} \hat{e}_z P_{circ} E_0^2
\end{equation}
Thus the presence of the CPGE at normal incidence allows
to conclude that the symmetry of the QW is not higher than  C$_s$.

\subsection*{\centerline{Experiment}}
The measurements were carried out
on $p$-type SiGe quantum well structures MBE-grown on (001)- and
(113)-oriented substrates. Two groups of (001)-grown samples were
fabricated in the following manner. One of the groups of samples had a
single quantum well of Si$_{0.75}$Ge$_{0.25}$ which was doped
with boron from one side only. The second group comprised ten stepped
quantum wells (Si$_{0.75}$Ge$_{0.25}$(4~nm)/
Si$_{0.55}$\,Ge$_{0.45}$(2.4~nm)), separated by 6~nm Si barriers. These
structures are of C$_{2v}$ point group symmetry which is  also
confirmed by the present experiment. Structures of the lower symmetry C$_s$
were (113)-grown with a Si/Si$_{0.75}$Ge$_{0.25}$(5~nm)/Si single QW
one-side boron doped. As a reference sample, a (001)-grown compositionally
symmetric (no step) and symmetrically boron doped multiple quantum well structure of
sixty Si$_{0.7}$Ge$_{0.3}$(3~nm) QWs has been used.

All these samples have free carrier densities of
about $8 \cdot 10^{11}$\,cm$^{-2}$ and were studied at
room temperature. For (001)-oriented
samples two pairs of ohmic point  contacts
were prepared corresponding to $x', y'$ parallel to  $\langle 110 \rangle$ and
$\langle 100 \rangle$-directions, respectively
(see inset Fig.~1).
Two additional pairs of contacts were formed in the center of the
sample edges with connecting lines along $x \parallel$
[1$\bar{1}$0] and $y \parallel$ [110].
For (113)-oriented samples two
pairs of contacts   were  centered along  opposite
sample edges pointing in the directions $x \parallel $ [1$\bar{1}$0]
and $y \parallel $ [33$\bar{2}$].

\begin{figure}[tb]
   \begin{center}
      \mbox{\epsfysize=4.8cm \epsfbox{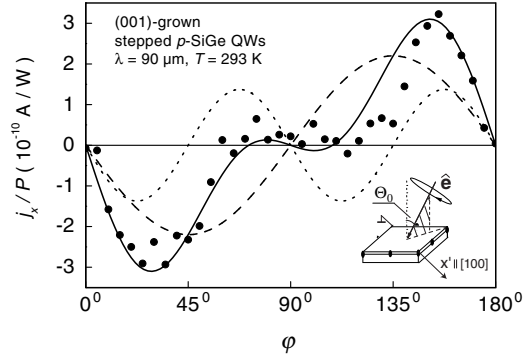}
}
\caption{Helicity dependence of photogalvanic current $j_x$
in (001)-grown and compositionally stepped
SiGe QWs normalized by the light power~$P$.
The data were obtained under oblique incidence $\Theta_0$= 30$^\circ$
of irradiation at $\lambda = 90\,\mu$m.
Broken and dotted lines show $j_x \propto \sin
2\varphi$ and $j_x \propto \sin 2\varphi \cdot \cos 2 \varphi$,
respectively. The full line is the sum of both.
  }
\end{center}
\end{figure}

A high power pulsed mid-infrared (MIR) TEA-CO$_2$ laser and a
far-infrared (FIR) NH$_3$-laser
have been used as radiation sources delivering 100\,ns
pulses with  radiation power  $P$ up to 100\,kW. Several
lines of the CO$_2$ laser between 9.2 and 10.6~$\mu$m
and of the NH$_3$-laser~\cite{PhysicaB99}  between
$\lambda$ = 76\,$\mu$m and 280\,$\mu$m
have been used
for excitation in the MIR and FIR range, respectively.
The MIR radiation induces direct optical transitions between heavy
hole and light hole subbands while the FIR radiation
causes indirect optical transitions in the lowest heavy-hole
subband.
The laser light polarization was modified from
linear to circular using for MIR  a Fresnel rhombus
and for FIR  quartz $\lambda/4$ plates. The
helicity of the incident light was varied
according to $P_{circ} = \sin{2
\varphi}$ where $\varphi$ is the angle between the initial
plane of linear polarization and the optical axis of the
$\lambda/4$ plate.

 With illumination by MIR radiation of the CO$_2$ laser
in (001)-oriented samples with asymmetric quantum wells, a
current signal proportional to the helicity $P_{circ}$ is
observed under oblique incidence as shown in Fig.~1.
The full line is $\propto\sin{2\varphi}$ ordinate scaled to the experimental data in agreement to Eq.~(\ref{jxjy}).
We note that the samples
were unbiased, thus the irradiated samples represent  current
sources. The magnitude of the current cannot be predicted as there is no microscopic theory for the tensor{\boldmath $\gamma$}.
The current follows the temporal structure of the
laser pulse intensity and changes sign if the circular
polarization is switched from left to right handed.
For $\langle 110 \rangle$ as well as $\langle 100 \rangle$ crystallographic
directions the photocurrent flows perpendicular to the wavevector of the
incident light. Therefore only a transverse CPGE was observed. It means that
effect  of the Dresselhaus {\boldmath$k$}-linear term
($\sigma_x k_x - \sigma_y k_y$)  is
negligible. The wavelength dependence of the photocurrent obtained between
9.2~$\mu$m and 10.6~$\mu$m corresponds to the spectral behaviour of direct
intersubband absorption between the lowest heavy-hole and light-hole
subbands~\cite{PRB02}.

In the FIR range a more complicated dependence of the
current as a function of helicity has been observed. In (001)-grown
asymmetric quantum wells as well as in (113)-grown samples
the observed dependence of the current on the phase angle $\varphi$
may be described by the sum of  two terms, one of them is
$\propto\sin 2\varphi$ and the other $\propto\sin2 \varphi \cdot \cos 2
\varphi$. In Fig.~2 experimental data and a fit to these
functions
are shown for a step bunched (001)-grown SiGe sample. The first term is due
to the CPGE and the second term is caused by the linear photogalvanic
effect~\cite{PRB02,APL00}.
For circularly polarized radiation the
$\sin 2 \varphi \cdot \cos 2\varphi$ term is equal to zero and the observed
current is due to the CPGE only.
\begin{figure}[tb]
   \begin{center}
      \mbox{\epsfysize=4.8cm \epsfbox{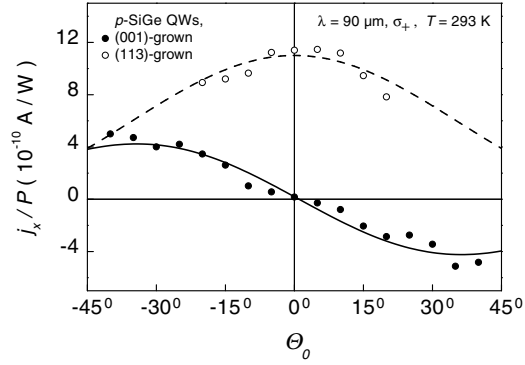}
}
\caption{
Photogalvanic current $j_x$ normalized by the light power
$P$ as a function of the angle of incidence.
The data were obtained for (001)-grown compositionally  stepped
and (113)-grown SiGe QWs irradiated
by right-handed circularly polarized light, $\sigma_+$,
at $\lambda = 90\,\mu$m.
The full line shows the result of calculation after the phenomenological theory.}
\end{center}
\end{figure}

With (001)-grown samples the signal vanishes at normal incidence, $\Theta_0$=0.
The variation of the angle of incidence
from positive to negative results in a change of direction of current
flow~( Fig.~3).
For (113)-grown samples the
current does not change its sign by the variation of
$\Theta_0$ and assumes a maximum at $\Theta_0$=0 (Fig.~3).
In symmetrically (001)-grown and symmetrically doped
SiGe quantum wells no photogalvanic current has been observed
in spite of the fact
that these samples, in order to increase their sensitivity,
contain substantially more quantum wells than the asymmetric structure
described above.
\subsection*{\centerline{Summary}}
We have shown that in asymmetric SiGe QWs the absorption of
circularly polarized radiation leads to spin photocurrents. This is
demonstrated by the observation  of
a helicity driven  current due to the CPGE.
Analysis of  the CPGE with respect to the symmetry of the
QWs allows to conclude
on  the spin-dependent {\boldmath$k$}-linear terms in the Hamiltonian.
Our results provide the important information that spin-related phenomena,
which so far have been considered to be specific for QW strucures based on
zinc-blende-structure materials, exist also in the SiGe QW systems.
\subsection*{\centerline{Acknowledgements}}
Financial support from the
DFG, the RFFI, the Russian Ministry of Science and the NATO linkage
program is gratefully acknowledged.

\end{document}